\documentclass[10pt,twocolumn,letterpaper]{article}

\usepackage{graphicx}
\usepackage{amsmath}
\usepackage{amssymb}
\usepackage{booktabs}
\usepackage{multirow}
\usepackage{lipsum}
\usepackage[table,xcdraw]{xcolor}

\usepackage[pagenumbers]{iccv} %

\usepackage{algorithm,algpseudocode}

\renewcommand{\paragraph}[1]{\vspace{2pt}\noindent{\bf #1}}

\definecolor{iccvblue}{rgb}{0.21,0.49,0.74}
\usepackage[pagebackref,breaklinks,colorlinks,allcolors=iccvblue]{hyperref}

\title{3D Gaussian Inverse Rendering with Approximated Global Illumination}

\author{\textbf{Zirui Wu}$^{1,2}$ \hspace{0.3em}
\textbf{Jianteng Chen}$^{2}$ \hspace{0.3em}
\textbf{Laijian Li}$^{2}$ \hspace{0.3em}
\textbf{Shaoteng Wu}$^{2}$ \hspace{0.3em}
\textbf{Zhikai Zhu}$^{2}$ \\
\textbf{Kang Xu}$^{2}$ \hspace{0.3em}
\textbf{Martin R. Oswald}$^{3}$ \hspace{0.3em}
\textbf{Jie Song}$^{1,4}$\\
$^{1}$ HKUST(GZ) \hspace{1em}
$^{2}$ NIO \hspace{1em}
$^{3}$ University of Amsterdam \hspace{1em}
$^{4}$ HKUST
}

\begin{document}

\newcommand{\indir}{\boldsymbol{\omega_i}}
\newcommand{\outdir}{\boldsymbol{\omega_o}}
\newcommand{\albedo}{\mathbf{a}}
\newcommand{\normal}{\mathbf{n}}
\newcommand{\halfvec}{\mathbf{h}}
\newcommand{\wzr}[1]{\textcolor{red}{(Zirui Memo) #1}}

\twocolumn[{%
\renewcommand\twocolumn[1][]{#1}%
\maketitle
\begin{center}
    \centering
    \vspace{-2em}
    \captionsetup{type=figure}
\includegraphics[width=\textwidth]{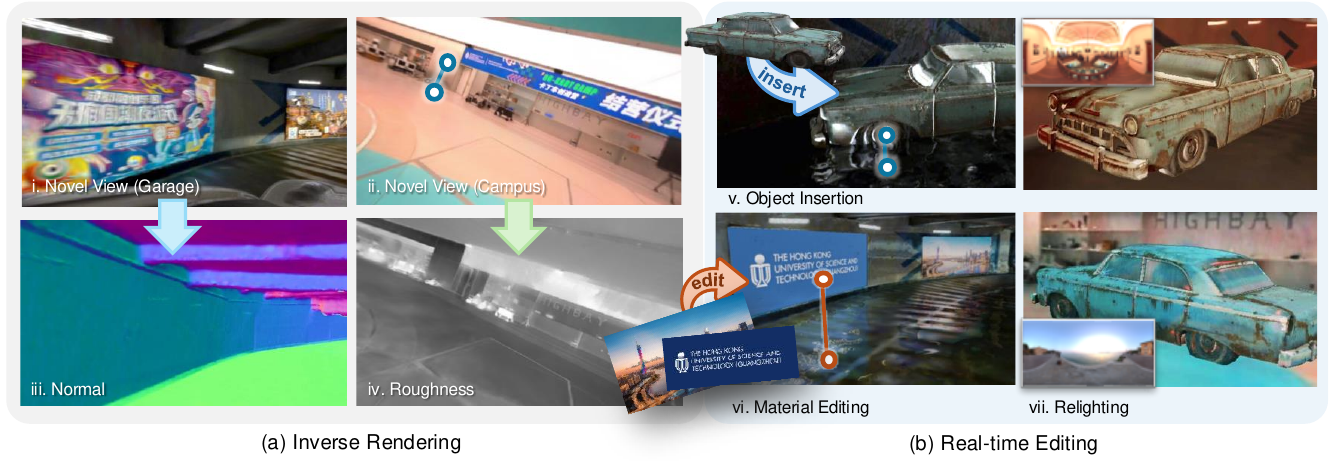}
    \captionof{figure}{\textbf{Overview}: (a) Our inverse rendering pipeline recovers geometry and material properties from 3D captures, visualized through normal (iii) and roughness (iv) maps.  The decomposition enables various editing capabilities: object insertion, material editing, and relighting. Our screen-space ray tracing technique ensures physically plausible reflections (visualized through corresponding point pairs).}
    \label{fig:teaser}
\end{center}%
}]

\begin{abstract}
3D Gaussian Splatting shows great potential in reconstructing photo-realistic 3D scenes. However, these methods typically bake illumination into their representations, limiting their use for physically-based rendering and scene editing. Although recent inverse rendering approaches aim to decompose scenes into material and lighting components, they often rely on simplifying assumptions that fail when editing.
We present a novel approach that enables efficient global illumination for 3D Gaussians Splatting through screen-space ray tracing. Our key insight is that a substantial amount of indirect light can be traced back to surfaces visible within the current view frustum. Leveraging this observation, we augment the direct shading computed by 3D Gaussians with Monte-Carlo screen-space ray-tracing to capture one-bounce indirect illumination. In this way, our method enables realistic global illumination without sacrificing the computational efficiency and editability benefits of 3D Gaussians. Through experiments, we show that the screen-space approximation we utilize allows for indirect illumination and supports real-time rendering and editing.
Code, data, and models will be made available at our project page: \href{https://wuzirui.github.io/gs-ssr}{https://wuzirui.github.io/gs-ssr}.
\end{abstract}

\section{Introduction}
Creating digital replicas of the physical world that support realistic simulations is a fundamental challenge in 3D computer vision and graphics. Physically-based rendering engines like Blender~\cite{community_blender3d_2018} enable the creation of virtual environments that adhere to real-world physics, allowing us to edit and simulate scenarios as if they truly existed in the physical world. For instance, autonomous driving researchers may need to insert cars into reconstructed scenes or modify lighting conditions (as in Fig.~\ref{fig:teaser}-b) to test their perception algorithms. With recent advances in neural rendering, especially Neural Radiance Fields ~\cite{mildenhall_nerfrepresenting_2020,muller_instantneural_2022,barron_mipnerfmultiscale_2021,zhu_nicerslamneural_2023} and 3D Gaussian Splattings~\cite{bernhard_3dgaussian_2023,yu_mipsplattingaliasfree_2023,ye_gaustudiomodular_2024,huang_2dgaussian_2024,yugay_gaussianslamphotorealistic_2023}, we can now reconstruct the 3D world and render photorealistic images using the reconstructed model. 

Among these advances, Gaussian Splatting is particularly promising for simulation environments, thanks to its discrete nature and efficient rendering capabilities. Representing scenes as individual Gaussian primitives naturally facilitates local editing operations, and its rasterization-based pipeline enables real-time performance. However, current methods primarily focus on reconstruction, where illumination and material properties are \textbf{baked} into their representation. This results in poor view extrapolation and, crucially for simulation purposes, the inability to edit material properties, scene geometry, and lighting conditions while maintaining physically plausible results (Fig.~\ref{fig:teaser}-b).

Recent inverse rendering techniques attempt to address these limitations by recovering explicit material and lighting properties~\cite{barron_shapeillumination_2020,liu_neroneural_2023}. These methods combine physically based rendering theory with Gaussian Splats~\cite{jiang_gaussianshader3d_2023,yang_specgaussiananisotropic_2024, zhihaoliang_gsir3d_2023}. However, most existing works~\cite{jiang_gaussianshader3d_2023, yang_specgaussiananisotropic_2024} only consider direct illumination - light arriving directly from sources like the sun or lamps. They typically approximate this using environment maps that assume all light originates from an infinite distance. This simplification breaks down in real-world scenes where indirect illumination - light bouncing between surfaces before reaching our eyes - plays a crucial role. As shown in Fig.~\ref{fig:teaser}-(ii), these indirect effects are ubiquitous, from reflections of billboards on the ground to subtle color bleeding between walls. Without modeling these complex light interactions, current approaches struggle to achieve realistic rendering results in such environments.

Incorporating global illumination effects into Gaussian Splatting is desirable but presents certain technical considerations. While Gaussian Splatting produces photorealistic results efficiently~\cite{bernhard_3dgaussian_2023}, its rasterization-based approach differs from ray tracing methods - instead of actively exploring points in 3D space, it passively receives points that are projected onto the canvas, making it challenging to track reflection paths and compute indirect illumination.

Previous works have attempted to address these challenges through various simplifying assumptions, but this often comes at the cost of real-time editability. For instance, GaussianShader~\cite{jiang_gaussianshader3d_2023} operates under an object-centric model, using a global environment map to represent all incident lights. While some studies~\cite{gao_relightable3d_2024,zhihaoliang_gsir3d_2023,guo_prtgsprecomputed_2024} tackle global illumination by pre-computing light-surface interactions into specific data structures, such as volumetric grids of indirect lights, these precomputed resources become invalid after any scene edits. This necessitates costly recomputation and ultimately limits real-time performance.

In this work, we bypass costly pre-computations and recomputations by utilizing rasterized G-buffers that store per-pixel geometric and material properties (Fig.~\ref{fig:teaser} iii,iv). 
Our key insight is that a significant portion of indirect illumination can be visible within the current view frustum (as shown by the connected point pairs in Fig.~\ref{fig:teaser}). We propose a novel approach that approximates global illumination through efficient screen-space ray tracing~\cite{mcguire_efficientgpu_2014}. 
Specifically, we first perform deferred shading to generate per-pixel G-buffers encoding surface geometry and material properties. Then, we execute a fast Monte-Carlo screen-space ray tracing step directly on these G-buffers to estimate the indirect illumination. Finally, we composite the resulting indirect illumination with direct shading to produce the final rendered image. 
This method is particularly effective in enclosed environments (e.g. underground garages) where most light-contributing and receiving surfaces are simultaneously visible. Unlike full global illumination methods that require complete scene geometry~\cite{moenne-loccoz_3dgaussian_2024,blanc_raygaussvolumetric_2024}, our screen-space ray tracing approach relies solely on the information available in the current frame, significantly reducing computation latency. Coupled with our customized CUDA kernel, this allows for real-time performance in both novel view synthesis and scene editing.

\noindent
In summary, we make the following \textbf{contributions}:
\begin{itemize}[noitemsep,topsep=0pt]
    \item We propose an inverse rendering framework for Gaussian Splattings that facilitates approximation of one-bounce global illumination by screen-space ray tracing while preserving real-time scene editability.
    \item Our framework accurately decomposes scene appearance into intrinsic surface properties and direct/indirect illumination components, enabling realistic edits to geometry, materials, and lighting conditions.
    \item Our approach delivers realistic global illumination effects while ensuring real-time performance. Our code will be made publicly available.
\end{itemize}

\section{Related Works}
\begin{figure*}[t]
    \centering
    \includegraphics[width=1\textwidth]{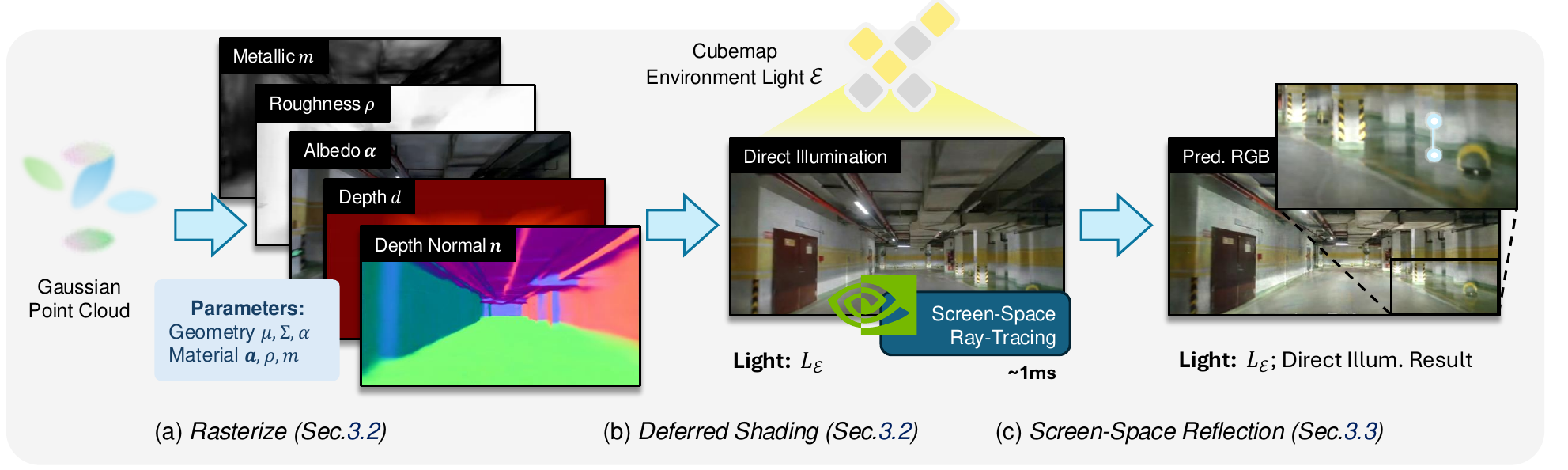}
    \caption{\textbf{Pipeline}: (a) Our method extends standard Gaussian splatting with material intrinsic properties. We first rasterize the Gaussian primitives into G-buffers containing both geometric and material properties of the current rendering frame; (b) We perform deferred shading on the alpha-composited G-buffers with direct environment lights from a learnable cubemap. (Sec.~\ref{sec:deferred-shading}); (c) We approximate one-bounce indirect lights through screen-space ray tracing and compose the final rendered RGB through Monte-Carlo integration. (Sec.\ref{sec:ssr}).}
    \label{fig:pipeline}
    \vspace{-2mm}
\end{figure*}
\paragraph{Inverse Rendering.}
Inverse rendering aims to decompose appearance into its intrinsic material and lighting components. It presents a more complex challenge than 3D reconstruction alone, as it is inherently ill-posed - multiple combinations of materials and lighting can produce identical appearances.

Traditional methods tackle this ambiguity through controlled lighting conditions~\cite{barron_shapeillumination_2020,lombardi_reflectanceillumination_2016,article,ackermann2015survey,zhang_ironinverse_2022}. FIPT~\cite{wu_factorizedinverse_2023} introduced a hybrid approach for global illumination with high physical accuracy, IRIS~\cite{lin_irisinverse_2025} successfully tackled the problem of HDR recovery from LDR inputs with physically-based rendering. Zhu et al.~\cite{zhu2022learning} demonstrated impressive results through Monte Carlo raytracing. The field has evolved with data-driven approaches that leverage large-scale models to learn intrinsic property priors~\cite{zeng_rgb-ximage_2024,ye_stablenormalreducing_2024}. Recent advances in neural representations have transformed inverse rendering approaches. Ref-NeRF~\cite{verbin_refnerfstructured_2022} introduced Integrated Directional Encoding (IDE) for view-dependent effects, NeILF~\cite{yao_neilfneural_2022,zhang_neilfinterreflectable_2023} enabled single-bounce global illumination via incident light field modeling. 

The emergence of 3D Gaussian Splatting (3DGS) has led to numerous innovations~\cite{shi_gir3d_2023,du_gsidillumination_2024,ye_3dgaussian_2024,wu_deferredgsdecoupled_2024}. GaussianShader~\cite{jiang_gaussianshader3d_2023} links material properties to Gaussians and models appearance through environmental lighting. GS-IR~\cite{zhihaoliang_gsir3d_2023} assumes static lighting and caches indirect illumination into a grid of light probes, which limits editability by fixing the indirect radiance in local light maps. Relightable3DGS~\cite{gao_relightable3d_2024} enables Monte-Carlo ray tracing through a bounding volume hierarchy (BVH) of Gaussians but requires costly BVH updates for any geometry modifications. PRTGS~\cite{guo_prtgsprecomputed_2024} encodes light transport in spherical harmonics (SH) parameters but still relies on expensive pre-computation during editing. In contrast, our method computes screen-space reflection dynamically, providing greater flexibility for modeling inter-reflections. Concurrently, GI-GS~\cite{chen_gigsglobal_2025} effectively leveraged screen-space techniques for efficient global illumination calculation. 

\paragraph{Neural Simulation.}
Neural simulation has become a vital technology that connects real-world perception with synthetic training environments for computer vision~\cite{guo_vid2avatar3d_2023} and robotics~\cite{wu_marsinstanceaware_2023,lou_robogsphysics_2024,zhong2025structured,10610694}. The Real2Sim pipeline facilitates cost-effective training of autonomous systems by creating digital twins of real environments. However, this is particularly challenging in driving scenarios due to the need to accurately model complex reflective surfaces, dynamic motion, sparse viewpoints, and various environmental factors, all while ensuring real-time editing capability.

Recent advances in driving simulation have primarily focused on modeling traffic agent behavior and scene dynamics through neural scene graphs~\cite{ost_neuralscene_2021,wu_marsinstanceaware_2023,yang_unisimneural_2023,chen_omnireomni_2024,zhou_hugsholistic_2024}. It was first used by Ost et al.~\cite{ost_neuralscene_2021}, who proposed separating the representation of dynamic vehicles from static backgrounds using distinct neural rendering models. This approach was further developed in UniSim~\cite{yang_unisimneural_2023} and MARS~\cite{walter_microfacetmodels_2007}, which incorporated efficient large-scale neural radiance fields~\cite{muller_instantneural_2022} to improve rendering quality and computational performance. OmniRe~\cite{chen_omnireomni_2024} extended these capabilities to handle general dynamic objects~\cite{yang_deformable3d_2023} and pedestrians, addressing a significant limitation in previous frameworks. 

However, these approaches share a common limitation: they typically encode lighting and material properties implicitly within their neural representations. This implicit encoding complicates the modification of scene illumination and the reuse of assets across different lighting conditions.
Our work addresses this gap by explicitly modeling material properties and enabling dynamic global illumination, supporting more flexible simulation environments.

\section{Methods}
As illustrated in Fig.~\ref{fig:pipeline}, our method extends Gaussian Splatting with physically-based materials and efficient screen-space global illumination. In Sec.~\ref{sec:prelim}, we review the fundamentals of physically-based rendering and the Gaussian Splatting framework. We then present our deferred shading pipeline in Sec.~\ref{sec:deferred-shading}, which efficiently processes the rasterized G-buffers to compute direct illumination. In Sec.~\ref{sec:ssr}, we detail our Monte-Carlo screen-space ray tracing technique that approximates indirect illumination. Finally, we describe our optimization framework in Sec.~\ref{sec:optimization}.

\subsection{Preliminaries}
\label{sec:prelim}
\paragraph{The Rendering Equation.}
The rendering equation~\cite{kajiya_renderingequation_1986} describes the light transport at any surface point $p$. The outgoing radiance $\mathbf{c}$ in direction $\outdir$ is the sum of reflected incident lights from all directions:
\begin{equation}
\label{eq:rendering}
    \mathbf{c}(\outdir) = \int_{\Omega^+} L_i(\indir)f(\indir,\outdir)(\indir\cdot\normal)\text{ d}\indir \, ,
\end{equation}
where $L_i(\indir)$ is the incident radiance from direction $\indir$ over the hemisphere $\Omega^+$, $f$ is the Bidirectional Reflectance Distribution Function (BRDF), and $\indir\cdot\normal$ accounts for the Lambert's cosine law. Unlike the original rendering equation~\cite{kajiya_renderingequation_1986}, we omit the emission term $\textbf{c}_{e}(\outdir)$. This design choice stems from our observation that emission terms can dominate the produced radiance during optimization, effectively suppressing proper material decomposition. Including such terms risks degrading our inverse rendering pipeline to behave like the original Gaussian Splatting, which lacks relighting capabilities.

\paragraph{3D Gaussian Rasterization.}
3D Gaussian Splatting (3DGS)~\cite{bernhard_3dgaussian_2023} represents a scene as a set of 3D Gaussian primitives that can be efficiently rasterized through alpha-blending. Each Gaussian is parameterized by its spatial position $\boldsymbol{\mu} \in \mathbb{R}^3$, covariance matrix $\boldsymbol{\Sigma} \in \mathbb{R}^{3\times3}$, and appearance attributes including opacity $\alpha$ and spherical harmonics coefficients for view-dependent color. The 3D Gaussian distribution is defined as:
\begin{equation}
    G(\mathbf{x}) = \exp(-\frac{1}{2}(\mathbf{x}-\boldsymbol{\mu})^\top\boldsymbol{\Sigma}^{-1}(\mathbf{x}-\boldsymbol{\mu}))\, .
\end{equation}
For efficient rendering, the covariance matrix is decomposed into rotation and scale components $\boldsymbol{\Sigma} = \mathbf{R}\mathbf{S}\mathbf{S}^\top\mathbf{R}^\top$, where $\mathbf{R}$ is a rotation matrix and $\mathbf{S}$ is a scaling matrix.

During rendering, these 3D Gaussians are projected to 2D screen space and alpha-blended in front-to-back order. The final pixel color $\mathbf{C}$ is computed as:
\begin{equation}
    \label{eq:alpha-blend}
    \mathbf{C} = \sum_{i=1}^N T_i\alpha_i\mathbf{c}_i\, ,
\end{equation}
where $T_i = \prod_{j=1}^{i-1}(1-\alpha_j)$ is the accumulated transmittance, $\alpha_i$ is the opacity, and $\mathbf{c}_i$ is the view-dependent color of the $i$-th Gaussian.

While this rasterization-based approach enables efficient rendering, it presents challenges for global illumination simulation. The key limitation stems from its \textbf{forward-only accumulation nature} -- each Gaussian can only receive illumination from directly visible light sources, making it difficult to model indirect bounces that require tracking light paths through multiple surfaces.

\subsection{Deferred Shading}
\label{sec:deferred-shading}
To enable physically-based rendering, we extend each Gaussian with material properties following Disney's principled BRDF~\cite{burley2012physically}. Specifically, each Gaussian is augmented with a diffuse albedo $\mathbf{a} \in [0,1]^3$, roughness $\rho \in [0,1]$, and metallic parameter $m \in [0,1]$. Following previous works~\cite{wu_deferredgsdecoupled_2024,ye_3dgaussian_2024}, we employ a deferred rendering process that separates the material composition and shading stages. We first alpha-composites the material parameters into 2D buffers (i.e. G-Buffers, see Fig.~\ref{fig:pipeline}-a) and then perform the shading computations on these buffers (Fig.~\ref{fig:pipeline}-b). Compared to the per-Gaussian shading process, this saves unnecessary computation on the Gaussians that are invisible to the current viewing frustum and prevents blending artifacts as stated in~\cite{wu_deferredgsdecoupled_2024}.

Similar to Eq.~\eqref{eq:alpha-blend}, the material properties are alpha-composited into G-buffers as:
\begin{equation}
\vspace{-0.2cm}
G_p = \sum_{i=1}^N T_i\alpha_i p_i \, ,
\label{eq:g-buffer}
\end{equation}
where $p \in \{\mathbf{a}, \rho, m, \gamma\}$ represents the material property.

Different from prior inverse rendering approaches~\cite{jiang_gaussianshader3d_2023,zhihaoliang_gsir3d_2023} that directly rasterize per-Gaussian normals, we compute surface normals from the rendered depth buffer (dubbed depth normal), which we found to be more robust in practice. Similar to 2DGS~\cite{huang_2dgaussian_2024}, the depth normal is computed through finite differences.
\begin{figure}[t]
    \centering
    \includegraphics[width=1\linewidth]{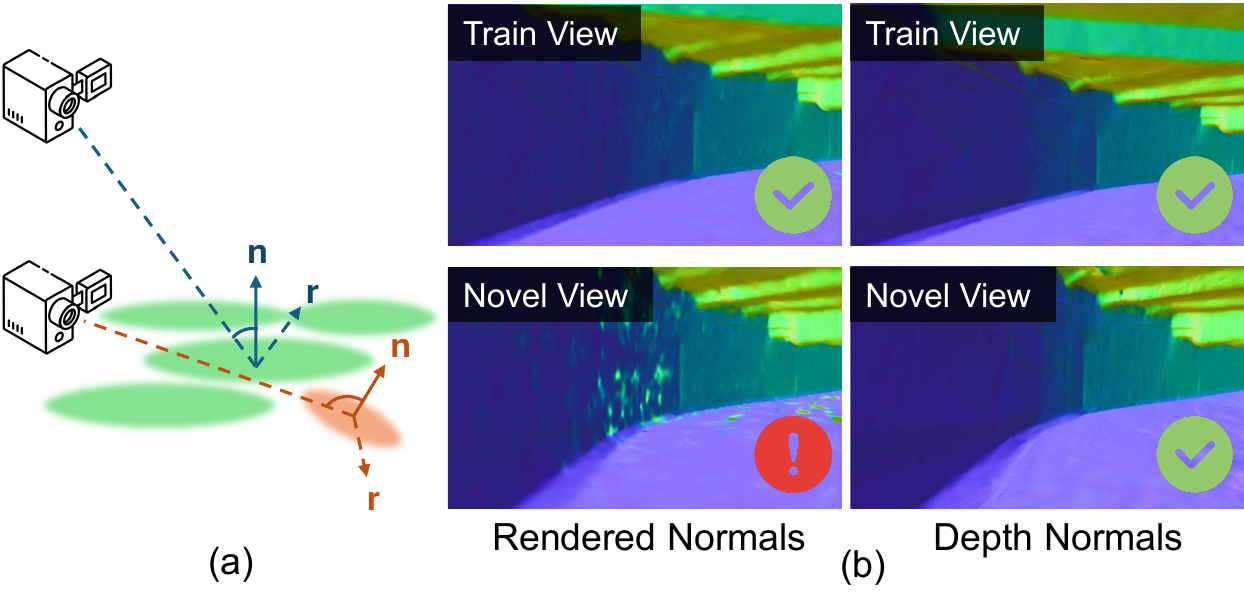}
    \caption{(a) While alpha-composited normals work well from the original camera view (top), they can produce incorrect estimates in novel views (bottom) when rays intersect with previously occluded surfaces. (b) Qualitative comparison showing that depth-based normals maintain consistency across both training and novel views, while rendered normals exhibit artifacts in novel views.}
    \label{fig:normal_gs_depth}
    \vspace{-1em}
\end{figure}

\paragraph{Per-Gaussian v.s. Depth Normal.}
While previous methods~\cite{jiang_gaussianshader3d_2023,zhihaoliang_gsir3d_2023} obtain surface normals through the alpha composition of per-Gaussian normals (Rendered Normals), we find that this can lead to inconsistent normal estimates at novel views (Fig.~\ref{fig:normal_gs_depth}-b, bottom left). This is because that the rendered normals only represent the orientation of \textbf{individual} Gaussians along each ray, without considering the underlying local geometry structure. In contrast, depth-based normal estimation leverages the spatial distribution of reconstructed surface points to compute geometrically meaningful normals (Fig.~\ref{fig:normal_gs_depth}-b, right column).
The superior robustness of depth normals is crucial for stable view synthesis, as accurate surface normals are essential for the shading process. Please find more details in the supplementary.

\paragraph{Physically-Based BRDF.}
We adopt a simplified Disney BRDF~\cite{burley2012physically} that models the surface interaction through a combination of diffuse and specular terms, parameterized by diffuse albedo $\albedo$, roughness $\rho$, and metallic $m$:
\begin{equation}
\label{eq:brdf}
f = \underbrace{\frac{1-m}{\pi}\albedo}_{f_d} + \underbrace{\frac{DFG}{4(\normal\cdot\indir)(\normal\cdot\outdir)}}_{f_s} \, ,
\end{equation}
where $D$, $F$, and $G$ represent the normal distribution, Fresnel, and geometry terms respectively. The equations for these terms are detailed in the supplementary materials.

Combining the BRDF with the rendering equation (Eq.~\eqref{eq:rendering}), the outgoing radiance of a shading point can be expressed as $\mathbf{c} = \mathbf{c}_d + \mathbf{c}_s$, where:
\begin{align}
    \mathbf{c}_d &= \frac{1-m}{\pi}\albedo\int_{\Omega^+}L_i(\indir)(\indir\cdot\normal)\text{ d}\indir \, ,\label{eq:diffuse_integral} \\
    \mathbf{c}_s &= \int_{\Omega^+}L_i(\indir)f_s(\indir\cdot\normal)\text{ d}\indir \, .\label{eq:specular_integral}
\end{align}
A direct solution to these integrals would require considering all possible light interactions between points in the scene, resulting in an algorithm with $\mathcal{O}(N^2)$ time complexity ($N$ being the number of points). This approach is computationally prohibitive in practice. To make physically-based rendering feasible with 3DGS, we employ the split-sum approximation~\cite{karis_realshading_2013} (detailed below), which allows us to consider only the dominant incident light direction. The approximations reduce the computational complexity to linear time while maintaining high rendering quality.

\paragraph{Shading.}
With the G-buffers providing the necessary information, we compute the first-pass shading that captures direct illumination from the environment. (Fig.~\ref{fig:pipeline}-b).

As solving the full integrals in Eq.~\eqref{eq:diffuse_integral} and \eqref{eq:specular_integral} is computationally prohibitive, we employ the split-sum approximation~\cite{karis_realshading_2013} that separates the ``sum of product" integral, where lighting and BRDF are entangled, into the product of two integrals. For the specular term $\textbf{c}_s$, this gives:
\begin{equation}
\label{eq:split-sum}
\mathbf{c}_s\!\approx\!\underbrace{\int_{\Omega^+}L_i(\indir)D(\rho, \mathbf{r})\text{ d}\indir}_\text{specular light integral}\cdot\underbrace{\int_{\Omega^+}f_s\cdot(\indir\cdot\normal)\text{ d}\indir}_{\text{BRDF integral } F_s} \, ,
\end{equation}
where $\textbf{r}$ is the reflection direction $\mathbf{r}$ calculated from the view direction $\mathbf{v}$ and surface normal $\mathbf{n}$ as $\mathbf{r} = 2(\mathbf{v}\cdot\mathbf{n})\mathbf{n} - \mathbf{v}$.

The BRDF integral can be pre-computed and stored in a 2D lookup table indexed by roughness $\rho$ and the cosine angle of $\textbf{n}\cdot\textbf{v}$. Detailed explanations are provided in the supplementary materials.

The specular light integral represents the color of the reflection lobe, whose direction follows $\textbf{r}$ and whose width is determined by the surface roughness $\rho$ - rougher surfaces receive a broader range of reflection. We implement this through a range query on the environment light using a learnable cubemap $\mathcal{E}$ with the NVDiffrast~\cite{laine_modularprimitives_2020} library. This process is expressed as:
\begin{equation}
L_\mathcal{E}(\mathbf{r}, G_\rho) = \mathrm{SampleEnvMap}(\mathbf{r}, \lambda, \mathcal{E}) \, ,
\end{equation}
where the mipmap level $\lambda$ is computed empirically as $\lambda = \log_2(G_\rho + 1)\lambda_{\max}$, with $\lambda_{\max}$ being the maximum mipmap level. This ensures proper integration over the specular lobe defined by the material roughness.

With these approximations, the specular term $\textbf{c}_s$ can be simplified to: 
\begin{equation}
    \textbf{c}_s=L_\mathcal{E}(\mathbf{r}, G_\rho)\cdot F_s,
\end{equation}
where $G_\rho$ is the alpha-composited roughness buffer as in Eq.~\eqref{eq:g-buffer}.
The diffuse color $\textbf{c}_d$ can be simplified similarly by querying the environment light map on the surface normal direction as
$\mathbf{c}_d = (1-G_m)G_\albedo \cdot L_\mathcal{E}(\mathbf{n}, G_\rho)$.
 The first-pass radiance $\mathbf{c}_1$ can then be obtained by combining the diffuse and specular components: $\mathbf{c}_1 = \mathbf{c}_d + \mathbf{c}_s$.
\begin{figure}[t]
    \centering
    \includegraphics[width=1\linewidth]{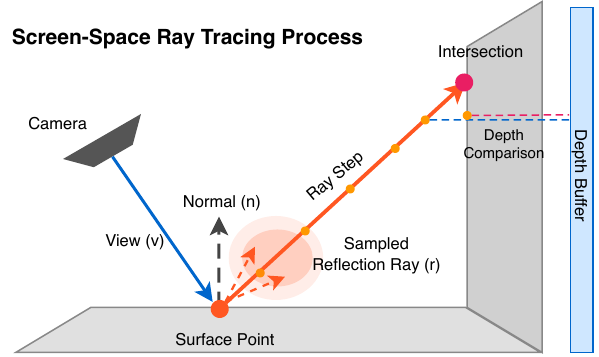}
    \caption{Screen-space ray tracing process. For each pixel, we march along the \textcolor[RGB]{237, 111, 69}{sampled reflection ray} step by step and compare the ray's depth with the scene \textcolor[RGB]{42, 100, 197}{depth buffer} at each step. An intersection is detected when the ray depth transitions from being in front of to behind the depth buffer values, indicating the ray has intersected with scene geometry.}
    \label{fig:ssr_concept}
    \vspace{-1em}
\end{figure}
\begin{table*}[t]
\centering
\resizebox{\textwidth}{!}{%
\begin{tabular}{clllllllllllllllllll}
\hline
 &
   &
  \multicolumn{3}{c}{\textbf{Garage-0}} &
  \multicolumn{3}{c}{\textbf{Garage-1}} &
  \multicolumn{3}{c}{\textbf{Garage-2}} &
  \multicolumn{3}{c}{\textbf{Garage-3}} &
  \multicolumn{3}{c}{\textbf{Campus-0}} &
  \multicolumn{3}{c}{\textbf{Campus-1}} \\
\multirow{-2}{*}{} &
   &
  PSNR* &
  SSIM* &
  LPIPS &
  PSNR* &
  SSIM* &
  LPIPS &
  PSNR* &
  SSIM* &
  LPIPS &
  PSNR* &
  SSIM* &
  LPIPS &
  PSNR &
  SSIM &
  LPIPS &
  PSNR &
  SSIM &
  LPIPS \\ \hline
 &
  PVG~\cite{chen_periodicvibration_2024} &
  \cellcolor[HTML]{FF8381}28.41 &
  \cellcolor[HTML]{FF8381}0.8415 &
  \cellcolor[HTML]{FFCE93}0.1645 &
  \cellcolor[HTML]{FF8381}26.99 &
  \cellcolor[HTML]{FF8381}0.6926 &
  \cellcolor[HTML]{FFCE93}0.2721 &
  \cellcolor[HTML]{FF8381}30.71 &
  \cellcolor[HTML]{FF8381}0.8486 &
  \cellcolor[HTML]{FFCE93}0.2355 &
  34.68 &
  0.9407 &
  \cellcolor[HTML]{FFCE93}0.1211 &
  28.36 &
  0.8408 &
  \cellcolor[HTML]{FFFC9E}0.2562 &
  \cellcolor[HTML]{FFFC9E}26.40 &
  \cellcolor[HTML]{FFFC9E}0.8792 &
  0.1881 \\
 &
  StreetGS~\cite{yan_streetgaussians_2024} &
  26.60 &
  0.8172 &
  0.1716 &
  25.52 &
  0.6245 &
  0.3395 &
  \cellcolor[HTML]{FFFC9E}27.77 &
  0.7811 &
  0.2844 &
  \cellcolor[HTML]{FF8381}36.39 &
  \cellcolor[HTML]{FF8381}0.9598 &
  0.1409 &
  \cellcolor[HTML]{FFFC9E}29.07 &
  \cellcolor[HTML]{FFFC9E}0.8627 &
  \cellcolor[HTML]{FFCE93}0.2018 &
  \cellcolor[HTML]{FF8381}26.54 &
  \cellcolor[HTML]{FFCE93}0.8888 &
  \cellcolor[HTML]{FF8381}0.1239 \\
 &
  OmniRe~\cite{chen_omnireomni_2024} &
  \cellcolor[HTML]{FFFC9E}26.83 &
  \cellcolor[HTML]{FFFC9E}0.8227 &
  \cellcolor[HTML]{FFFC9E}0.1712 &
  \cellcolor[HTML]{FFFC9E}25.72 &
  \cellcolor[HTML]{FFFC9E}0.6291 &
  \cellcolor[HTML]{FFFC9E}0.3370 &
  \cellcolor[HTML]{FFCE93}28.25 &
  \cellcolor[HTML]{FFCE93}0.7965 &
  \cellcolor[HTML]{FFFC9E}0.2818 &
  \cellcolor[HTML]{FFFC9E}36.04 &
  0.9586 &
  \cellcolor[HTML]{FFFC9E}0.1402 &
  \cellcolor[HTML]{FFCE93}29.38 &
  \cellcolor[HTML]{FFCE93}0.8631 &
  \cellcolor[HTML]{FFCE93}0.2018 &
  \cellcolor[HTML]{FFCE93}26.53 &
  \cellcolor[HTML]{FF8381}0.8889 &
  \cellcolor[HTML]{FFCE93}0.1241 \\
 &
  GShader~\cite{jiang_gaussianshader3d_2023} &
  25.02 &
  0.7343 &
  0.2837 &
  23.70 &
  0.5334 &
  0.5060 &
  24.56 &
  0.6571 &
  0.4259 &
  33.88 &
  0.9029 &
  0.1530 &
  24.57 &
  0.7821 &
  0.4463 &
  25.04 &
  0.8646 &
  0.1755 \\
\multirow{-5}{*}{NVS} &
  Ours &
  \cellcolor[HTML]{FFCE93}28.14 &
  \cellcolor[HTML]{FFCE93}0.8374 &
  \cellcolor[HTML]{FF8381}0.1587 &
  \cellcolor[HTML]{FFCE93}25.99 &
  \cellcolor[HTML]{FFCE93}0.6609 &
  \cellcolor[HTML]{FF8381}0.1694 &
  27.72 &
  0.7816 &
  \cellcolor[HTML]{FF8381}0.2295 &
  \cellcolor[HTML]{FFCE93}36.23 &
  \cellcolor[HTML]{FFCE93}0.9583 &
  \cellcolor[HTML]{FF8381}0.0424 &
  \cellcolor[HTML]{FF8381}29.66 &
  \cellcolor[HTML]{FF8381}0.8646 &
  \cellcolor[HTML]{FF8381}0.1812 &
  26.37 &
  0.8704 &
  \cellcolor[HTML]{FFFC9E}0.1389 \\ \hline
 &
  PVG~\cite{chen_periodicvibration_2024} &
  \cellcolor[HTML]{FF8381}28.90 &
  \cellcolor[HTML]{FFCE93}0.7986 &
  \cellcolor[HTML]{FFCE93}0.1621 &
  \cellcolor[HTML]{FF8381}28.89 &
  \cellcolor[HTML]{FFCE93}0.7987 &
  0.2940 &
  \cellcolor[HTML]{FF8381}32.20 &
  \cellcolor[HTML]{FF8381}0.8842 &
  0.2291 &
  \cellcolor[HTML]{FF8381}38.47 &
  \cellcolor[HTML]{FFFC9E}0.9631 &
  \cellcolor[HTML]{FFCE93}0.1367 &
  30.19 &
  0.8749 &
  0.2329 &
  28.15 &
  0.9017 &
  0.1750 \\
 &
  StreetGS~\cite{yan_streetgaussians_2024} &
  \cellcolor[HTML]{FFFC9E}27.78 &
  \cellcolor[HTML]{FFFC9E}0.8537 &
  0.1655 &
  \cellcolor[HTML]{FFFC9E}27.46 &
  \cellcolor[HTML]{FFFC9E}0.7764 &
  \cellcolor[HTML]{FFCE93}0.2655 &
  28.92 &
  0.8229 &
  0.2776 &
  \cellcolor[HTML]{FFFC9E}38.23 &
  \cellcolor[HTML]{FFCE93}0.9646 &
  0.1499 &
  \cellcolor[HTML]{FFCE93}30.88 &
  \cellcolor[HTML]{FFFC9E}0.8941 &
  \cellcolor[HTML]{FFCE93}0.1919 &
  \cellcolor[HTML]{FFCE93}29.12 &
  \cellcolor[HTML]{FFCE93}0.9223 &
  \cellcolor[HTML]{FF8381}0.1126 \\
 &
  OmniRe~\cite{chen_omnireomni_2024} &
  27.30 &
  0.7746 &
  \cellcolor[HTML]{FFFC9E}0.1635 &
  27.30 &
  0.7746 &
  \cellcolor[HTML]{FFFC9E}0.2689 &
  \cellcolor[HTML]{FFCE93}29.63 &
  \cellcolor[HTML]{FFCE93}0.8436 &
  \cellcolor[HTML]{FFFC9E}0.2747 &
  \cellcolor[HTML]{FFCE93}38.43 &
  \cellcolor[HTML]{FF8381}0.9691 &
  \cellcolor[HTML]{FFFC9E}0.1413 &
  \cellcolor[HTML]{FFFC9E}30.86 &
  \cellcolor[HTML]{FFCE93}0.8960 &
  \cellcolor[HTML]{FFFC9E}0.1925 &
  \cellcolor[HTML]{FF8381}29.15 &
  \cellcolor[HTML]{FF8381}0.9224 &
  \cellcolor[HTML]{FF8381}0.1126 \\
 &
  GShader~\cite{jiang_gaussianshader3d_2023} &
  26.57 &
  0.7965 &
  0.272 &
  23.89 &
  0.5608 &
  0.5025 &
  27.01 &
  0.7706 &
  0.2867 &
  36.54 &
  0.9606 &
  0.1792 &
  25.03 &
  0.7920 &
  0.4388 &
  26.90 &
  0.8957 &
  \cellcolor[HTML]{FFFC9E}0.1657 \\
\multirow{-5}{*}{Recon} &
  Ours &
  \cellcolor[HTML]{FFCE93}28.82 &
  \cellcolor[HTML]{FF8381}0.8720 &
  \cellcolor[HTML]{FF8381}0.1348 &
  \cellcolor[HTML]{FFCE93}28.47 &
  \cellcolor[HTML]{FF8381}0.8232 &
  \cellcolor[HTML]{FF8381}0.1591 &
  \cellcolor[HTML]{FFFC9E}29.16 &
  \cellcolor[HTML]{FFFC9E}0.8288 &
  \cellcolor[HTML]{FF8381}0.1945 &
  36.97 &
  0.9562 &
  \cellcolor[HTML]{FF8381}0.0458 &
  \cellcolor[HTML]{FF8381}32.02 &
  \cellcolor[HTML]{FF8381}0.9066 &
  \cellcolor[HTML]{FF8381}0.1610 &
  \cellcolor[HTML]{FFFC9E}29.07 &
  \cellcolor[HTML]{FFFC9E}0.9115 &
  \cellcolor[HTML]{FFCE93}0.1205 \\ \hline
\end{tabular}%
}
\caption{Quantitative evaluation results. * For PSNR and SSIM computation, we exclude the region that contains the ego car in the image.}
\label{tab:quantitative}
\end{table*}

\subsection{Screen-space Ray Tracing}
\label{sec:ssr}
While the first-pass shading captures direct illumination, it fails to account for indirect bounces that are crucial for realistic rendering, particularly in indoor environments. We propose to incorporate an efficient screen-space ray tracing approach that approximates one-bounce indirect illumination by leveraging the information available in the G-buffers~\cite{mcguire_efficientgpu_2014} (Fig.~\ref{fig:pipeline}-c).

We illustrate the screen-space tracing process in Fig.~\ref{fig:ssr_concept}. For each pixel, we trace reflection rays in screen space to find potential indirect light contributions (shown as corresponding point pairs in Fig.~\ref{fig:pipeline}-c). Given a pixel's world position $\mathbf{p}$ (un-projected from the depth buffer), view direction $\mathbf{v}$, and normal $\mathbf{n}$, we generate a reflection ray as $\textbf{r}(t)$ with $t$ being the marched distance.

To detect intersections, we project the ray into screen space and march along it in fixed steps. At each step $i$, we compare the ray's depth $z_\text{ray}$ with the scene depth $z_\text{scene}$ from the depth buffer:
\begin{equation}
\Delta z_i = z_\text{ray}(t_i) - z_\text{scene}\big(\mathrm{proj}\big(\mathbf{r}(t_i)\big)\big) \, ,
\end{equation}
where $\text{proj}$ denotes perspective projection. An intersection is detected when $\Delta z_i$ changes from negative to positive, indicating the ray has passed through a surface. We denote the UV coordinate of the intersection point as $p^*$.

In practice, we sample $N_s=8$ reflection rays from the reflection lobe and trace the reflection colors $\{\textbf{c}_{1,i}\}_{N_s}$ in the screen space. Then, the sampled colors are combined using Monte-Carlo integration:
\begin{equation}
    \textbf{c}_s'=\frac{1}{N_s} \sum_{i=1}^{N_s} \frac{f_s \textbf{c}_{1,i}(\omega_i \cdot n)}{ p_{GGX}(\omega_i)},
\end{equation}

where $\omega_i$ is the $i$-th reflection direction sampled from the GGX~\cite{walter_microfacetmodels_2007} distribution $p_{GGX}$.

The final rendered color is obtained by combining the diffuse component with the Monte Carlo integrated specular reflection, followed by tone mapping and gamma correction to convert the result into the standard RGB color space.

\subsection{Optimization}
\label{sec:optimization}
Our optimization objective combines several losses to ensure high-quality reconstruction of both geometry and material properties. The total loss is defined as:
\begin{equation}
\mathcal{L}_\text{total} = \mathcal{L}_\text{rgb} + \lambda_\text{o}\mathcal{L}_\text{opacity} + \lambda_\text{n}\mathcal{L}_\text{n} + \lambda_\text{reg}\mathcal{L}_\text{reg} \, ,
\end{equation}
where $\mathcal{L}_\text{rgb}$ is the standard RGB reconstruction loss, $\mathcal{L}_\text{opacity}$ encourages the Gaussians to fully cover the images, $\mathcal{L}_\text{n}$ is a depth-normal consistency loss following~\cite{huang_2dgaussian_2024}, and $\mathcal{L}_\text{reg}$ is a total variance loss applied on the G-buffers. Details on the loss functions are provided in the supplementary.

\section{Experiments}

\paragraph{Baselines.}
We compare our method with 3 lines of baseline methods:
1) Dynamic view synthesis algorithm: We select Periodic Vibration Gaussian (PVG)~\cite{chen_periodicvibration_2024}, which models appearance changes through temporal-dependent Gaussians.
2) Driving Simulators: We compare against Street Gaussians (StreetGS)~\cite{yan_streetgaussians_2024}, which augments 3D Gaussians with dynamic spherical harmonics for appearance changes. We also compare with OmniRe~\cite{chen_omnireomni_2024}, which constructs dynamic neural scene graphs using multiple types of dynamic nodes to model various kinds of actors in driving scenes. 
3) Inverse rendering algorithm: We include GaussianShader (GShader)~\cite{jiang_gaussianshader3d_2023} as our primary baseline for inverse rendering comparison, as it also builds upon Gaussian Splatting but only uses cubemap lighting.

To ensure fair comparisons, all methods are initialized with identical point clouds derived from LiDAR captures. We maintain the original hyperparameters as specified in each method's paper. For PVG~\cite{chen_periodicvibration_2024} and StreetGS~\cite{yan_streetgaussians_2024}, we utilize the implementations provided in the OmniRe repository~\cite{chen_omnireomni_2024}. GShader~\cite{jiang_gaussianshader3d_2023} is integrated directly into our evaluation pipeline through code migration to ensure consistent evaluation conditions.

\paragraph{Dataset.}
Our evaluation uses a self-collected dataset comprising 4 sequences from underground garages and 2 campus scenes. Garage sequences are captured using a multi-sensor setup: three synchronized RGB cameras operating at 30 FPS, accompanied by LiDAR scans for geometric reference. Campus scenes are captured with a hand-held scanner with calibrated LiDAR and images~\cite{liu_omnicolorglobal_2024}. These environments present challenging lighting conditions with multiple light sources, significant indirect illumination, and various surface materials.

We conduct quantitative experiments under two protocols established by prior works~\cite{ost_neuralscene_2021,chen_omnireomni_2024,wu_marsinstanceaware_2023}:
1) Scene Reconstruction: Using all frames for both training and testing to evaluate the method's reconstruction fidelity;
2) Novel View Synthesis: Using 50\% of frames for training and the remainder for testing, selecting test frames uniformly across the sequence to assess generalization.
\subsection{Quantitative Evaluations}

\paragraph{Metrics.}
Due to the presence of ego vehicle parts in our captured images, we modify all baseline methods to exclude those regions in the reconstruction losses during training. Each method is trained for 30,000 steps. We evaluate performance using standard image quality metrics: PSNR, SSIM, and LPIPS~\cite{zhang_unreasonableeffectiveness_2018}. For PSNR and SSIM calculations, we exclude regions containing the ego vehicle to focus on scene reconstruction quality.

\paragraph{Results.} Table~\ref{tab:quantitative} shows the quantitative comparison results. PVG~\cite{chen_periodicvibration_2024} (first row) serves as an effective upper bound in our quantitative evaluations due to its unique approach to modeling time-dependent Gaussians. While it cannot extrapolate views with physically plausible reflections, it achieves superior reconstruction and view interpolation quality by encoding light-surface interactions as temporal features. This makes it particularly effective at reproducing the training sequence, albeit without the physical interpretability that our method provides.

Beyond the comparison with PVG, our method is on par with other 3D reconstruction baselines (row 2 and 3) in both settings and consistently outperforms the inverse rendering baseline GShader~\cite{jiang_gaussianshader3d_2023} (fourth row), which lacks enough capacity to model indirect lighting effects.
\begin{figure}[t]
    \centering
    \includegraphics[width=\linewidth]{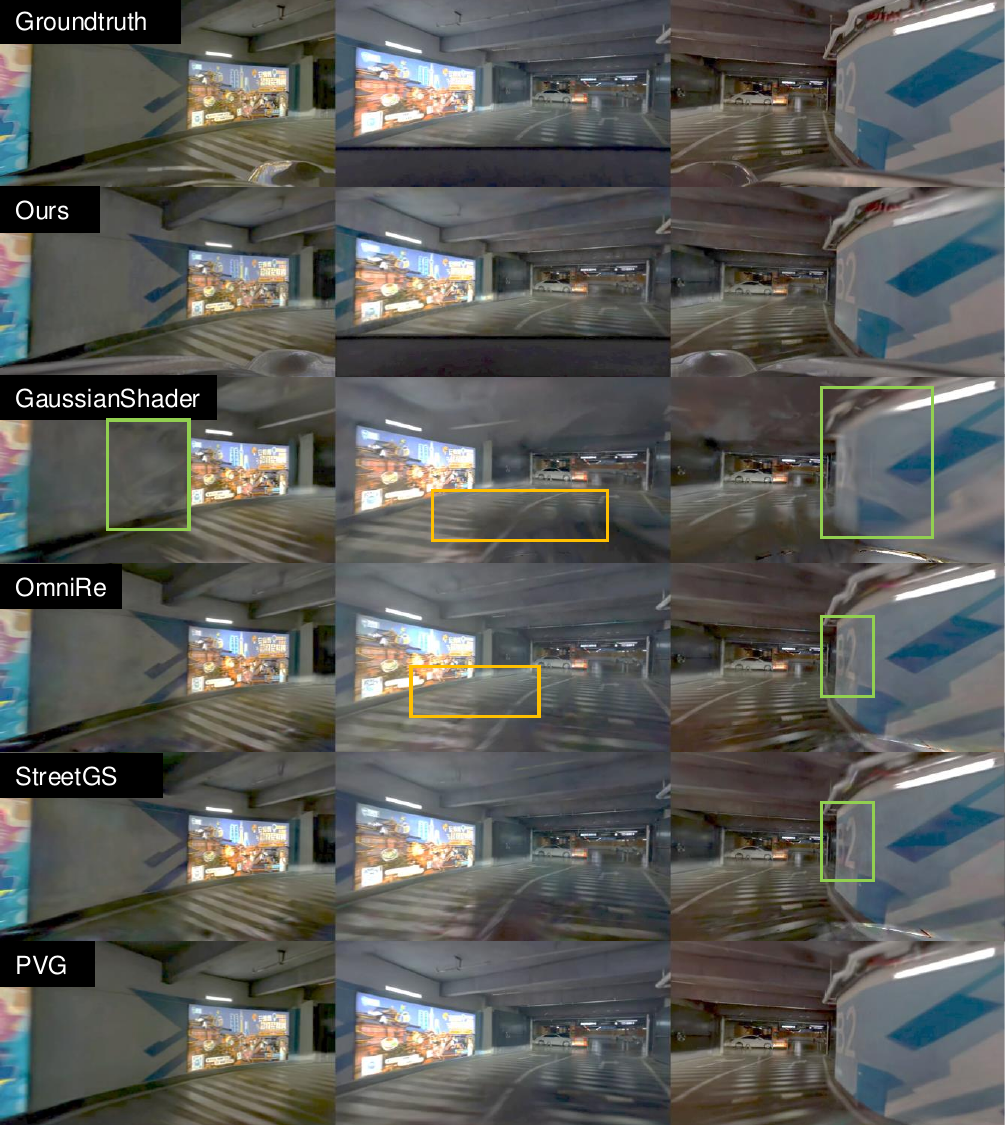}
    \caption{Qualitative comparisons. We show the rendering results for each method. The regions highlighted in \textcolor[RGB]{245, 194, 66}{yellow} boxes show reflections on the ground where baseline methods struggle to capture dramatically dynamic indirect illumination effects; \textcolor[RGB]{94, 129, 63}{Green} boxes show blurry artifacts.}
    \label{fig:qualitative}
    \vspace{-1em}
\end{figure}

\begin{table}[t]
\centering
\resizebox{\columnwidth}{!}{%
\begin{tabular}{ccccccc}
\hline
    & PVG~\cite{chen_periodicvibration_2024} & StreetGS~\cite{yan_streetgaussians_2024} & OmniRe~\cite{chen_omnireomni_2024} & GShader~\cite{jiang_gaussianshader3d_2023} & Ours w/o SSR & Ours \\ \hline
FPS &  72   &    69  &  71    &      74          &    38  &    37    \\ \hline
\end{tabular}%
}
\caption{Real-time rendering performance comparison. FPS is measured using an NVIDIA 4090 at 960$\times$540 resolution.}
\label{tab:fps}
\vspace{-1em}
\end{table}

\paragraph{Rendering speed.}
We evaluate the real-time rendering performance of our method against the baselines. As shown in Table~\ref{tab:fps}, reconstruction-based methods like PVG~\cite{chen_periodicvibration_2024}, StreetGS~\cite{yan_streetgaussians_2024}, OmniRe~\cite{chen_omnireomni_2024}, and GShader~\cite{jiang_gaussianshader3d_2023} achieve 69-74 FPS. Our method maintains 37 FPS while computing physically-based shading and global illumination effects. The small overhead between our full method and the variant without screen-space reflections (38 vs 37 FPS) demonstrates that our ray marching implementation adds minimal computational cost.

\subsection{Qualitative Evaluation}
We demonstrate our method's capabilities through several visual experiments:

\paragraph{Reconstruction.} As shown in Fig.~\ref{fig:qualitative}, our method achieves faithful reconstruction quality compared to baseline approaches, particularly in handling reflective surfaces and indirect illumination effects. The most notable differences appear in the reflective floor regions (highlighted in yellow boxes) where GShader~\cite{jiang_gaussianshader3d_2023} exhibits significant artifacts due to its limited modeling of global illumination. Standard reconstruction methods like OmniRe~\cite{chen_omnireomni_2024} and StreetGS~\cite{yan_streetgaussians_2024} also struggle with these challenging lighting scenarios, producing blurry or inconsistent reflections (highlighted in green boxes). In contrast, our physically-based shading model successfully captures the complex interplay of light, accurately reproducing both the billboard reflections on the floor and the subtle ambient lighting throughout the parking garage. PVG~\cite{chen_periodicvibration_2024} achieves comparable visual quality but does so through temporal encoding rather than explicit physical modeling.
\begin{figure}[t]
    \centering
    \includegraphics[width=\linewidth]{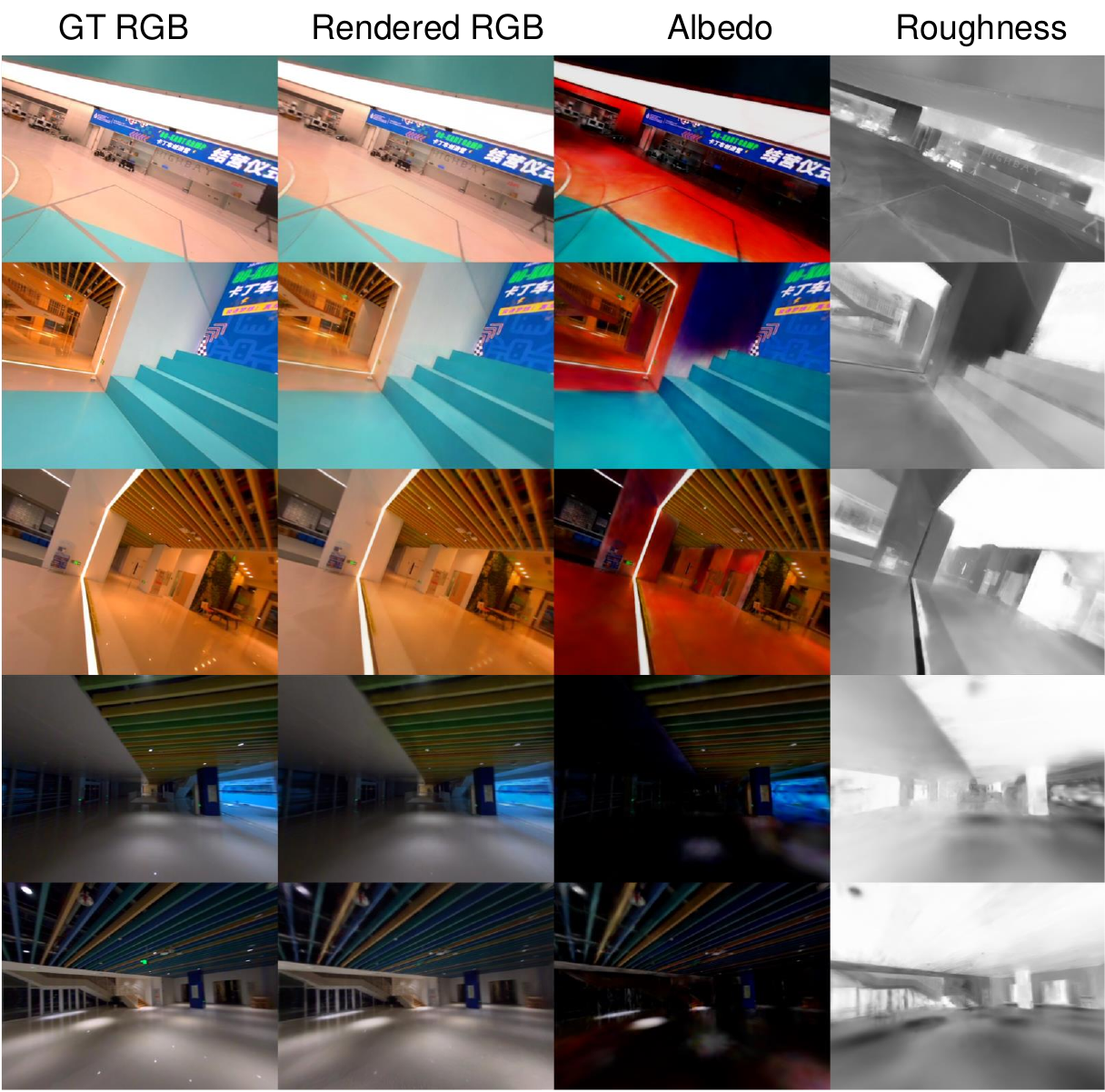}
    \caption{Decomposition results. We show our decomposed intrinsic channels. The contrast ratio of the consistency channels is amplified for visualization.}
    \label{fig:decomposition}
    \vspace{-1.5em}
\end{figure}

\begin{figure}[t]
    \centering
    \vspace{-1em}
    \includegraphics[width=.98\linewidth]{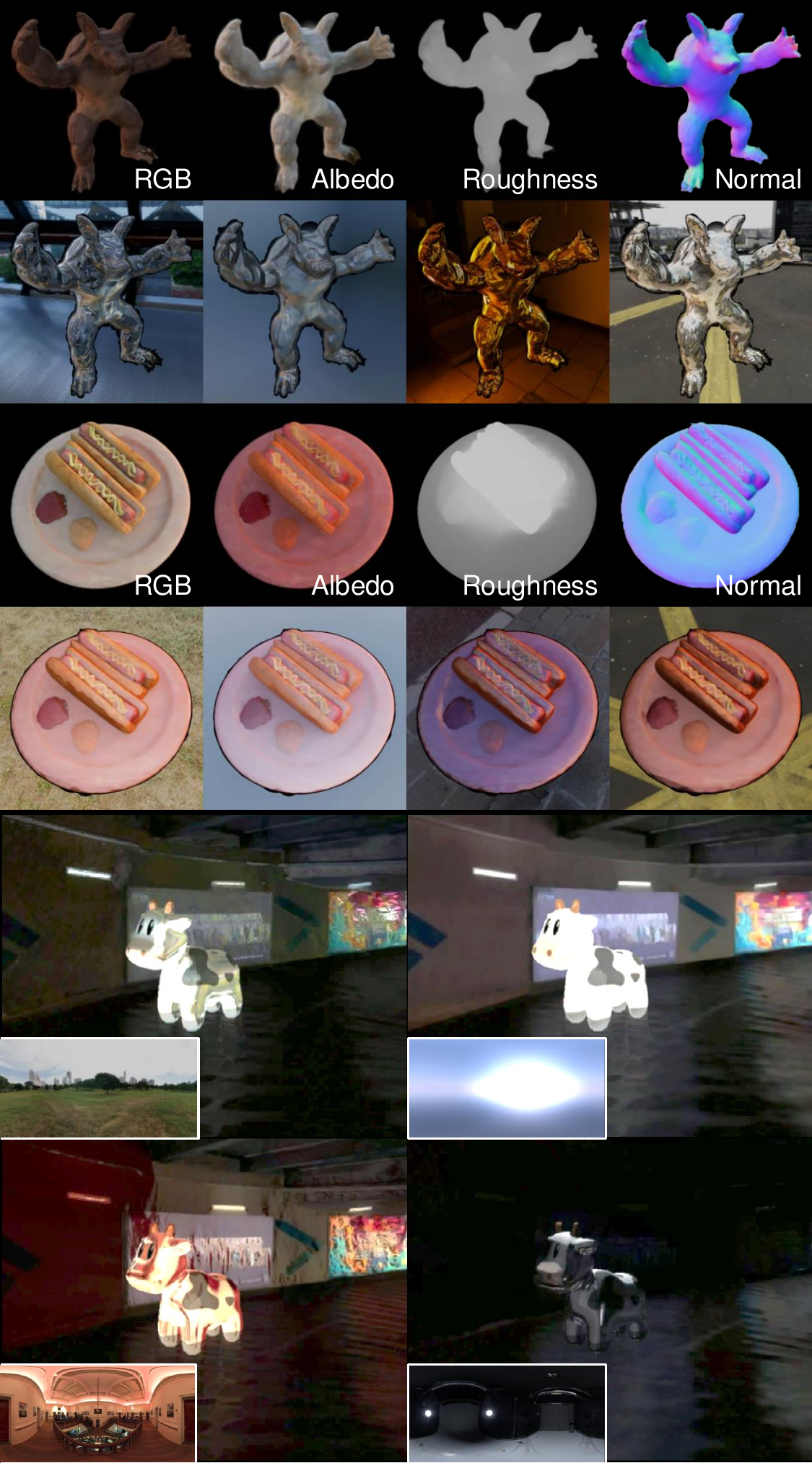}
    \caption{Top: Intrinsic channels and editing results (relighting, material editing) from NeRF-Synthetic dataset~\cite{mildenhall_nerfdark_2021}. Bottom: Relighted garage scene with inserted object.}
    \label{fig:editing}
    \vspace{-1.5em}
\end{figure}

\paragraph{Decomposition.}
Our method achieves convincing decomposition of scene appearance into its constituent components. As shown in Fig.~\ref{fig:decomposition}, we can separate the complex visual effects in indoor scenes into intrinsic components. The albedo map reveals the intrinsic surface colors without the influence of illumination. We also show decomposition results of small objects from TensoIR~\cite{jin_tensoirtensorial_2024} dataset in Fig.~\ref{fig:editing}.

\paragraph{Relighting.} As demonstrated in Fig.~\ref{fig:editing}, our method successfully handles various lighting conditions by incorporating HDRI environment maps from online sources~\cite{humus,hdrihaven,jin_tensoirtensorial_2024}.

\paragraph{Object Insertion.} We demonstrate the seamless integration of external 3D assets into the reconstructed scene. The process begins by rendering the inserted object into G-buffers using NVDiffrast~\cite{laine_modularprimitives_2020}. These buffers are then composited with the scene's G-buffers through depth-based comparison. As shown in Fig.~\ref{fig:editing}, this approach enables the inserted object to receive proper illumination while contributing to indirect lighting effects, evidenced by the realistic reflections on the floor.

\paragraph{Material Editing.} Our method enables direct manipulation of material properties, including roughness and metallic parameters (Fig.~\ref{fig:editing}). By adjusting these physically-based parameters, we can achieve a range of surface appearances from diffuse to highly specular, while maintaining consistent interactions with both direct and indirect illumination.

\begin{figure}
    \centering
    \includegraphics[width=1\linewidth]{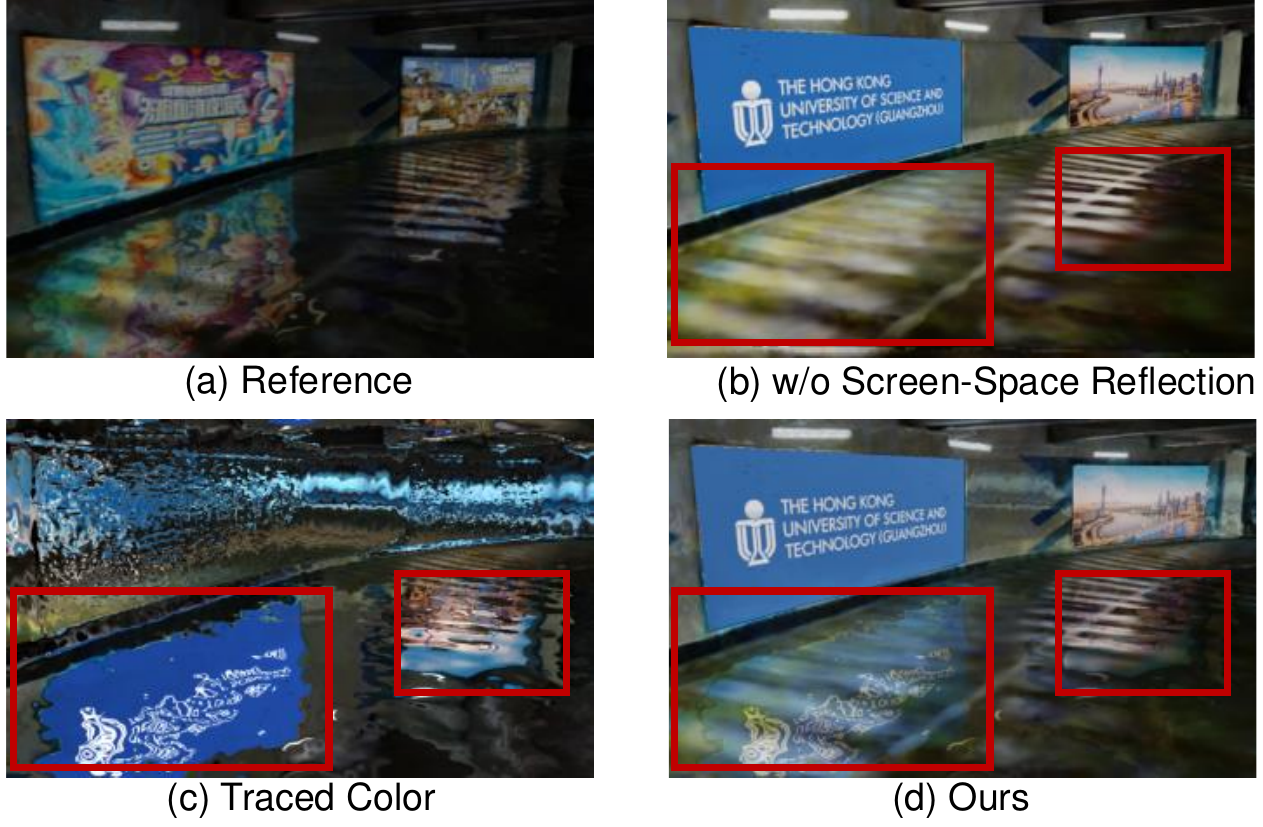}
    \caption{Effect of screen-space reflections during editing. (a) Reference image. (b) Without modeling screen-space reflections, the floor reflections remain unchanged after editing since they are baked into the albedo. (c) Visualization of the screen-space traced lights. (d) Our method reflects the edited billboard content. We set roughness to a relatively low value to exaggerate the effect.}
    \label{fig:ablation_ssr}
    \vspace{-1.5em}
\end{figure}

Our decomposed representation also allows for intuitive editing of surface colors. As demonstrated in Fig.~\ref{fig:ablation_ssr}, we can easily modify scene content by editing the albedo channel. Our method automatically updates all lighting interactions to maintain physical consistency, with floor reflections dynamically adapting to the new content. This showcases how our approach preserves the natural interaction between edited materials and global illumination.

\paragraph{Effect of Screen-Space Reflections.} Fig.~\ref{fig:ablation_ssr} also serves to demonstrate the importance of our screen-space reflection technique. Without it, reflections become baked into the ground floor albedo thus showing noticeable artifacts at novel views (b). Our method properly handles indirect illumination through ray tracing, ensuring reflections update accurately with edited content (d).

\section{Conclusion}

We have developed a novel inverse rendering framework for Gaussian Splatting that enables real-time editing with global illumination effects. By combining screen-space ray tracing with Gaussian representations and a learned consistency parameter, our method achieves physically plausible rendering and interactive performance. It allows for scene decomposition into editable components, facilitating operations like object insertion, material editing, and relighting in complex indoor environments. Experimental results confirm the effectiveness of our approach and realistic screen-space reflections.
However, our method has limitations. The non-differentiable nature of screen-space ray marching can impact optimization stability, suggesting a need for numerical gradient techniques. Our method's performance decreases in outdoor settings with distant light interactions, indicating a need for a hybrid approach to managing varying scales of light transport.

{
    \small
    \bibliographystyle{ieee_fullname}
    \bibliography{wuzirui_bibtex,additional}
}

\appendix
\section{Acknowledgement}
We thank Ziyi Yang and Haozhe Lou for the fruitful discussions, Bonan Liu for the help on high-quality data collection, Jiajun Jiang and Handi Yin for proofreading. 

\section{Microfacet BRDF}
We use the GGX-Trowbridge-Reitz distribution~\cite{walter_microfacetmodels_2007} for the normal distribution function:
\begin{equation}
\label{eq:ndf}
    D(\halfvec;\rho,\normal) = \frac{\rho^2}{\pi((\halfvec\cdot\normal)^2(\rho^2-1)+1)^2},
\end{equation}
where $\halfvec = (\indir + \outdir)/|\indir + \outdir|$ denote the half vector between incident and outgoing directions.

The Fresnel term models view-dependent reflectance using Schlick's approximation:
\begin{equation}
\label{eq:fresnel}
    F(\outdir, \halfvec, \albedo, m) = F_0 + (1 - F_0)(1-(\outdir\cdot\halfvec)^5),
\end{equation}
where $F_0 = \text{lerp}(0.04, \albedo, m)$ interpolates between dielectric and metallic surfaces.

The geometry term models microfacet shadowing and masking:
\begin{equation}
\label{eq:geometry}
    G(\indir, \outdir, \normal, \rho) = G_\text{GGX}(\indir\cdot\normal)G_\text{GGX}(\outdir\cdot\normal),
\end{equation}
where $G_\text{GGX}(z) = \frac{2z}{z+\sqrt{\rho^2+(1-\rho^2)z^2}}$.

\section{Split-sum Approximation Details}
\label{sec:split_sum_supp}

The split-sum approximation separates the rendering integral into a BRDF term that can be efficiently pre-computed. Using the Schlick approximation, the Fresnel term is simplified to depend on only the albedo $\textbf{a}$ and the half angle $\theta$. 

By applying Schlick's approximation, the BRDF integral can be split into two terms:
\begin{align}
&\int_{\Omega^+}f_r(p, \omega_i, \omega_o) \cos \theta_i d\omega_i \nonumber \\
\approx&\ \textbf{a} \int_{\Omega^+} \frac{f_r}{F} (1-(1-\cos \theta_i)^5) \cos \theta_i d\omega_i \\
&+\int_{\Omega^+} \frac{f_r}{F} (1-\cos \theta_i)^5 \cos \theta_i d\omega_i\nonumber
\end{align}

These two integrals are pre-computed through Monte-Carlo integration and stored as 2D lookup tables parameterized by surface roughness $\rho$ and viewing angle $\cos \theta$. The lookup table encodes the BRDF response in its color channels - the red channel stores the first integral results while the green channel contains the second integral. This pre-computation approach enables efficient runtime evaluation of the BRDF while preserving visual accuracy.

\section{Screen-space Ray Tracing Algorithm}
We detail the screen-space ray tracing algorithm at Algorithm~\ref{alg:ssr}. The algorithm begins at a pixel location $\mathbf{p}_{uv}$, computes the reflection direction $\mathbf{r}$ based on the view direction $\mathbf{v}$ and surface normal $\mathbf{n}$, and then marches along this direction in fixed steps $\triangle s$. At each step, it projects the world-space position $\mathbf{p}_{w}$ back to screen space and compares the ray's depth $d_\text{ray}$ with the scene depth $d_\text{scene}$ from the depth buffer. When these depths match within a threshold, we have found a reflection point.

\begin{algorithm}[t]
\caption{Screen Space Ray Tracing}
\label{alg:ssr}
\begin{algorithmic}[1]
\Function{TraceReflection}{\text{pixel}, \text{cameraPos}, initialStep, maxRayLength, \text{threshold}}
    \State $\mathbf{o} \gets \text{WorldPosition}(\text{pixel})$
    \State $\mathbf{n} \gets \text{SurfaceNormal}(\text{pixel})$
    \State $\mathbf{v} \gets \text{normalize}(\mathbf{o} - \text{cameraPos})$
    
    \State // Compute reflection direction
    \State $\mathbf{r} \gets \mathbf{v} - 2(\mathbf{v} \cdot \mathbf{n})\mathbf{n}$
    
    \State // Ray marching parameters
    \State $\triangle s \gets \text{initialStep}$
    \State $\text{maxRayLength}$
    \State $\text{bestHit} \gets \text{null}$
    
    \State // March ray through screen space
    \For{$dist \gets 0$ \textbf{to} $\text{maxRayLength}$ \textbf{step} $\triangle s$}
        \State $\mathbf{p}_{w} \gets \mathbf{o} + \mathbf{r} \times dist$
        \State $\mathbf{p}_{uv} \gets \text{ProjectToScreen}(\mathbf{p}_{w})$
        
        \If{$\text{IsOffScreen}(\mathbf{p}_{uv})$}
            \State \textbf{break}
        \EndIf
        
        \State $d_\text{scene} \gets \text{SampleDepthBuffer}(\mathbf{p}_{uv})$
        \State $d_\text{ray} \gets \mathbf{p}_{w}.z$
        
        \If{$|d_\text{ray} - d_\text{scene}| < \text{threshold}$}
            \State $\text{bestHit} \gets \mathbf{p}_{uv}$
            \State \textbf{break}
        \EndIf
    \EndFor
    
    \Return $\text{bestHit}$
\EndFunction
\end{algorithmic}
\end{algorithm}

\begin{figure*}[t]
    \centering
    \includegraphics[width=1\textwidth]{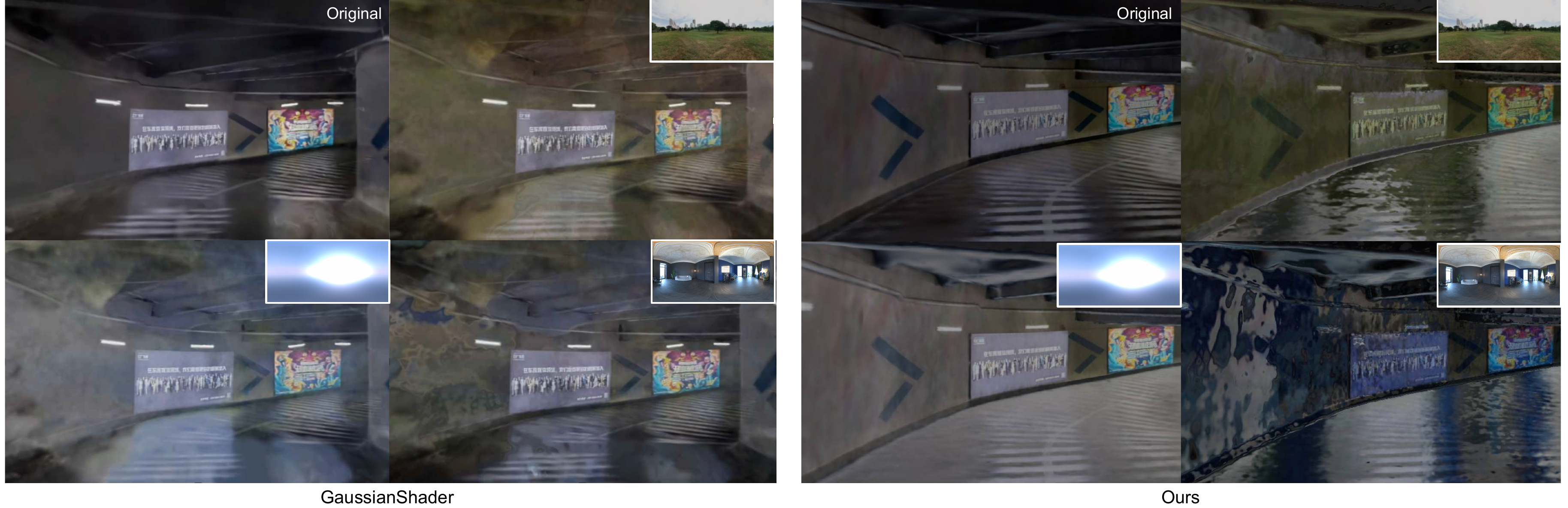}
    \caption{Qualitative comparisons of our method against GaussianShader~\cite{jiang_gaussianshader3d_2023} on relighting effects. GaussianShader overfits the training sequence with false geometry, thus producing significant artifacts while relighting.}
    \label{fig:relight_compare}
\end{figure*}

\section{Loss Functions}
Following the original 3D Gaussian Splatting framework, we employ RGB reconstruction loss and SSIM loss for appearance supervision $\mathcal{L}_\text{rgb} =$
\begin{align}
\lambda_\text{1} \|\mathbf{C} - \mathbf{C}_\text{gt}\|_1 + (1-\lambda_\text{1})(1 - \text{SSIM}(\mathbf{C}, \mathbf{C}_\text{gt})).
\end{align}

To ensure complete surface coverage, we introduce an opacity loss that encourages Gaussians to densely populate valid regions:
\begin{equation}
\mathcal{L}_\text{opacity} = \|1 - \sum_{i=1}^N T_i\alpha_i\|_2^2.
\end{equation}

To ensure a robust geometry optimization, we leverage an off-the-shelf monocular normal estimator~\cite{ye_stablenormalreducing_2024} to provide reliable guidance:
\begin{equation}
\mathcal{L}_\text{n} = 1 - (G_\mathbf{n} \cdot \mathbf{N}_\text{mono}).
\end{equation}
where $G_\mathbf{n}$ is our computed normal buffer and $\mathbf{N}_\text{mono}$ is the estimated ground truth normal.

To promote spatially coherent materials, we apply smoothness regularization on the rendered G-buffers as:
\begin{equation}
\mathcal{L}_\text{reg} = \sum_{p \in \{\mathbf{a}, \rho, m, \gamma\}} \|\nabla G_p\|_1.
\end{equation}

\section{Additional Experiment Details}
\paragraph{Implementation Details.}
Our data comes from vehicle-mounted cameras, where parts of each image contain the capturing vehicle. We handle this by learning an RGBA mask overlay on the rendered images to exclude vehicle-occupied regions from model training. The final image $\mathbf{I}$ is composited as:
\begin{equation}
\mathbf{I} = \alpha\mathbf{M} + (1-\alpha)\mathbf{R}
\end{equation}
where $\mathbf{M}$ is the learned RGB mask, $\alpha$ is its opacity, and $\mathbf{R}$ is the rendered image. Rather than simply masking out these regions during training, learning an RGBA mask enables us to synthesize novel views that maintain the appearance of being captured from the same vehicle.

To account for lighting variations across sequences, we designate every 10th frame as a keyframe with a separate environment light map, using linear interpolation between neighboring keyframes for intermediate frames. For scenes with moving objects, we follow OmniRe~\cite{chen_omnireomni_2024} by representing dynamic traffic elements as separate Gaussian sets. Since initializing with dense LiDAR points, we disable the adaptive density control mechanism from 3DGS~\cite{bernhard_3dgaussian_2023}. Our implementation builds upon OmniRe's official codebase, with all experiments conducted on a single NVIDIA RTX 4090 GPU.

\begin{figure}[t]
    \centering
    \includegraphics[width=1\linewidth]{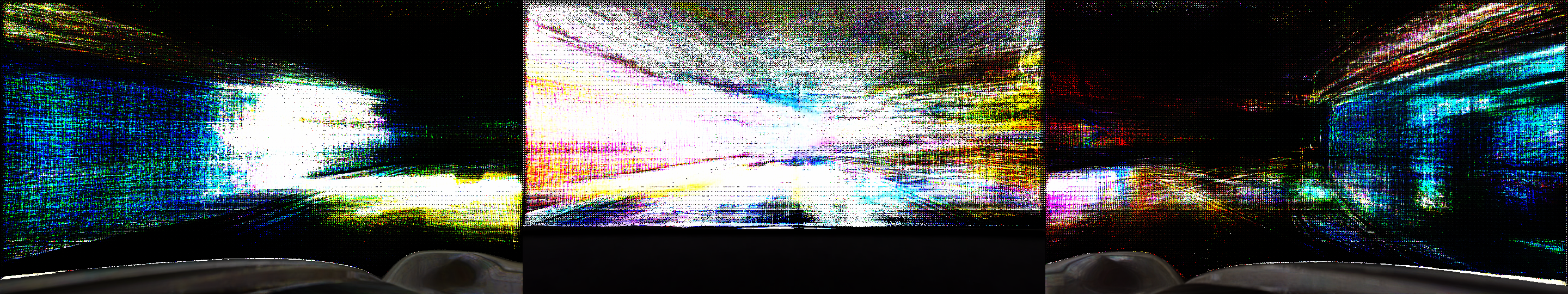}
    \caption{The learned ego car mask for 3 cameras. Only the RGB channels are displayed here without the alpha channel. The mask effectively removes the ego vehicle regions from training while preserving the surrounding scene information, enabling novel view synthesis that maintains the appearance of being captured from the same vehicle setup.}
    \label{fig:ego_rgb}
\end{figure}

\begin{figure}[t]
    \centering
    \includegraphics[width=1\linewidth]{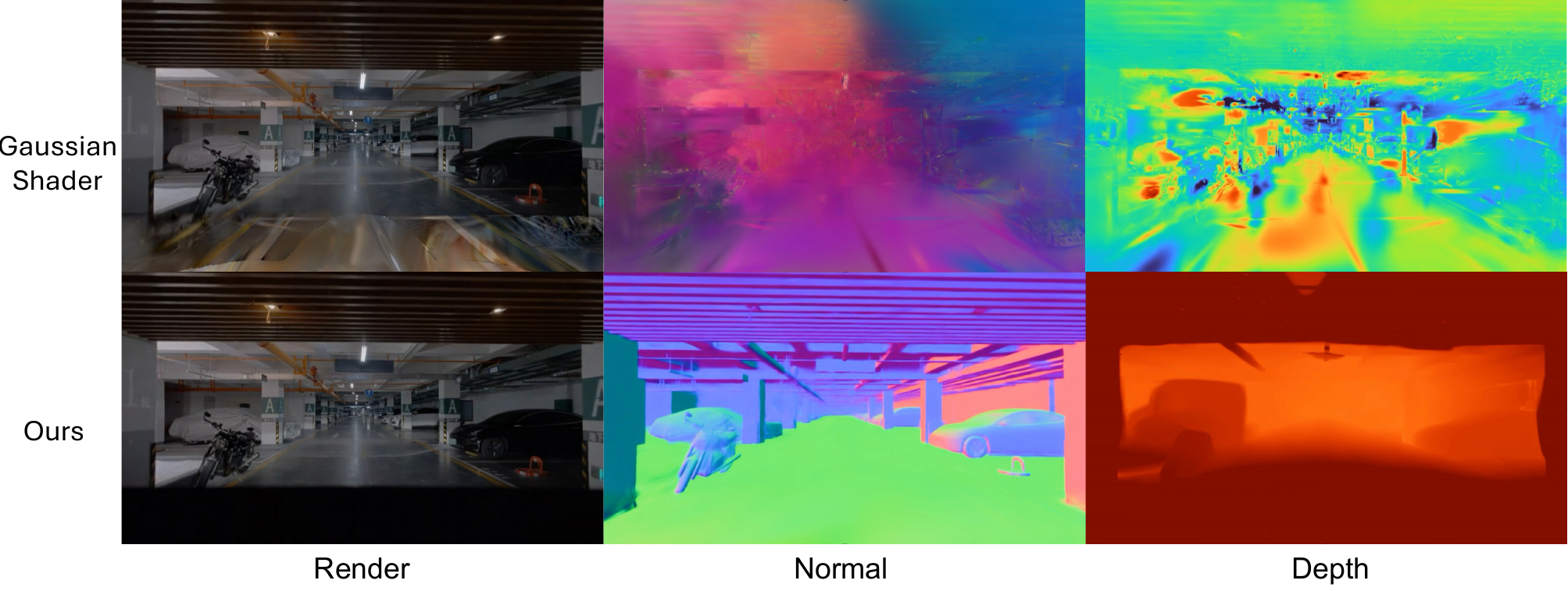}
    \caption{Comparison of scene decomposition results. Our method produces more coherent geometry representations compared to GaussianShader~\cite{jiang_gaussianshader3d_2023}.}
    \label{fig:decomp_compare}
\end{figure}

\paragraph{Decomposition Analysis.}
While both methods achieve similar rendering quality, our approach produces more physically meaningful scene decomposition. As shown in Fig.~\ref{fig:decomp_compare}, the normal maps from GaussianShader~\cite{jiang_gaussianshader3d_2023} exhibit noise and inconsistencies, particularly visible in the ceiling structure and ground plane. In contrast, our method generates clean, consistent normal maps that accurately capture the geometric structure of the scene. The depth maps further demonstrate this difference - our method produces sharp, well-defined depth boundaries that align with the actual scene geometry, while GaussianShader's depth estimates appear more ambiguous, especially at object boundaries. This improved geometric representation is crucial for downstream editing tasks, as it provides a more reliable basis for operations like relighting and material editing.

\paragraph{Relighting Quality.}
The benefits of our improved geometric representation become evident in relighting tasks. As demonstrated in Fig.~\ref{fig:relight_compare}, we can successfully relight the scene using different environment maps while maintaining scene coherence. This versatility in relighting is directly enabled by our method's accurate geometry estimation and physically-based rendering approach, which existing methods fell shorts at.

\begin{figure}
    \centering
    \includegraphics[width=1\linewidth]{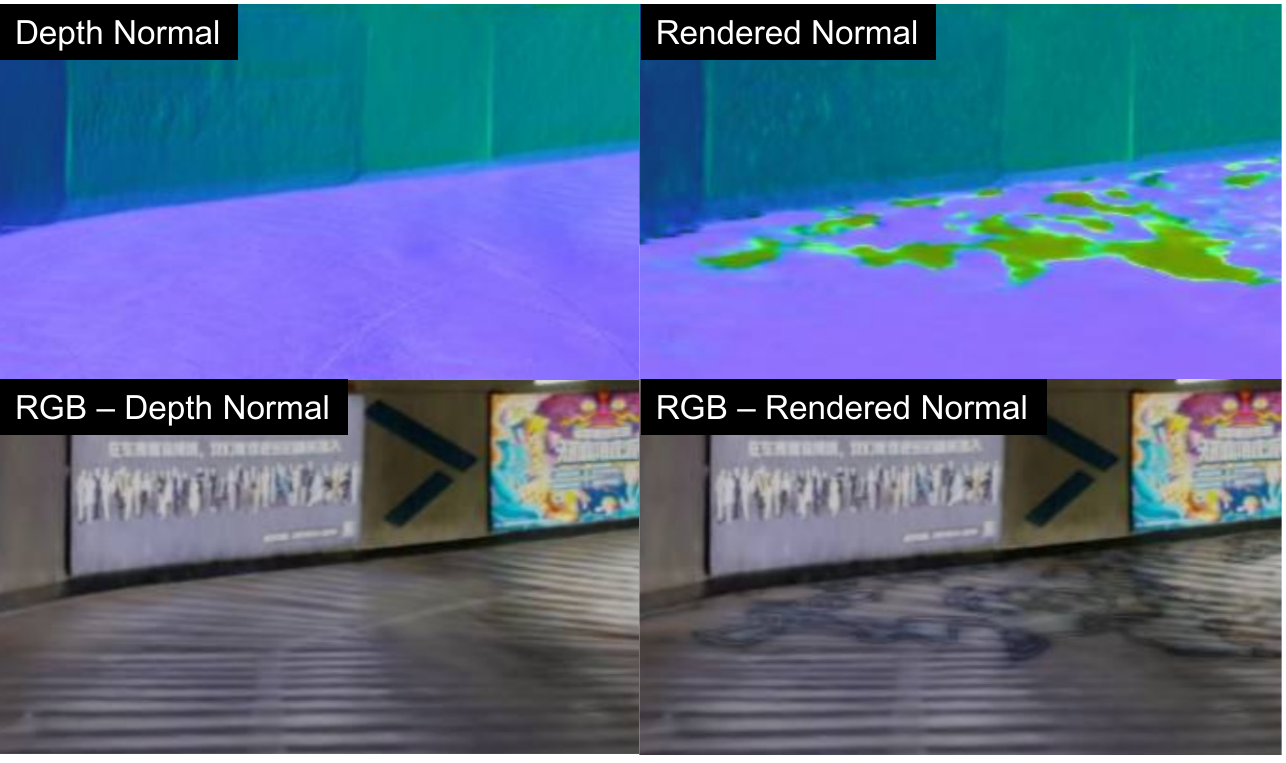}
    \caption{Comparison between depth-based and rendered normal estimation approaches. Top: normal visualization shows that rendered normals exhibit noisy artifacts on the floor, while depth normals maintain smooth and consistent surface orientation. Bottom: the resulting RGB renders demonstrate how these normal differences affect the final appearance, with rendered normals producing inconsistent specular reflections on the floor.}
    \label{fig:ablation_normal}
\end{figure}

\paragraph{Depth Normal vs. Per-Gaussian Normal.} 
We evaluate the trade-off between different normal estimation approaches. As shown in Fig.~\ref{fig:ablation_normal}, the depth-based normal estimation produces more stable and physically plausible results compared to per-Gaussian rendered normals. The normal maps (top row) reveal that rendered normals introduce high-frequency noise on planar surfaces like the floor, while depth normals maintain smooth and coherent surface orientation. This difference is reflected in the final renderings (bottom row), where depth normals enable more consistent specular reflections. Although quantitative metrics (Tab.~\ref{tab:ablation_normal}) show a small decrease in reconstruction accuracy with depth normals, we find this trade-off acceptable given the significant improvement in physical plausibility and editing stability.

\begin{table}[ht]
\centering
\begin{tabular}{llll}
\hline
                        & PSNR* & SSIM*  & LPIPS  \\ \hline
w/ per-Gaussian normals & 27.97 & 0.8457 & 0.1412 \\
w/ depth normals        & 27.71 & 0.8386 & 0.1479 \\ \hline
\end{tabular}%
\caption{Quantitative comparison of normal estimation methods.
* indicates that we excluded the region containing the ego car.}
\label{tab:ablation_normal}
\end{table}

\end{document}